\documentclass[12pt,preprint]{aastex}

\usepackage{epsfig}

\slugcomment{Not to appear in Nonlearned J., 45.}

\shorttitle{Empirical Counts of Active Galaxies and QSOs}
\shortauthors{P. Andreani et al.}

\begin{document}

\title{How many active galaxies and QSOs will future Space Missions detect?}

\author{Paola Andreani}\affil{INAF - Osservatorio Astronomico
di Padova, vicolo dell'Osservatorio 5, I-35122 Padova, Italy} \email{andreani@pd.astro.it}

\author{Luigi Spinoglio}\affil{Istituto di Fisica dello Spazio Interplanetario, CNR
via Fosso del Cavaliere 100, I-00133 - Roma, Italy}
\email{luigi@ifsi.rm.cnr.it}

 \and

\author{Matthew A. Malkan}\affil{Physics and Astronomy Department,
University of California, Los Angeles, CA, USA}
\email{malkan@astro.ucla.edu}

\begin{abstract}
Averaged spectral energy distributions (SEDs)
of {\it active and starburst galaxies} from the 12\,$\mu$m sample in the Local
Universe and {\it quasars}, from an optically selected sample at a mean redshift
$\langle z\rangle$=0.7 are built from optical, near-IR and far-IR (IRAS and ISO)
photometric observations.
These SEDs are then used to predict at various redshifts the number of Seyfert
type 1 and type 2, starburst, normal galaxies, and quasars, that will be detected
by future Space Missions dedicated to far-infrared and submillimeter astronomy,
like SIRTF and Herschel.
These predictions are then compared with the expected capabilities and
detection limits of future deep far-IR surveys. Possible ways to identify
AGN candidates on far-IR colour-colour plots for follow-up
observations are then explored.

\end{abstract}


\section{Introduction}
\label{Introduction}

Surveys in the far-infrared (FIR) and sub-millimetre (submm) region
from future space missions will provide crucial 
information on star formation and galaxy formation and evolution.
If AGN were in place roughly at the same
time as galaxies, it is natural to expect that these surveys will
shed light on the AGN role in the early evolution of
galaxies \cite[]{SR98,Gra01}.
\hfill\break
Both AGN and star formation
are characterized by a strong FIR emission both in the continuum and
in molecular and atomic lines, because of the dusty and gaseous
environment where they release their energy.
An early dust-enshrouded phase of AGN is expected, but
is missed in current surveys based on other techniques.
The optical AGN phase could occur late in the process of QSO
formation, corresponding to the
exhaustion of the gaseous-dusty shroud inside which the host
galaxy formed.

The original 12$\mu$m active galaxy sample \cite[]{SM89,RMS}
provided a complete and largely unbiased sample of
local active galaxies.  It is relatively free from
the strong selection effects of
optical-UV surveys, which are biased in favor of type 1 AGN, and
far-IR surveys, which are biased in favor of dusty type 2 AGN and
starburst galaxies.
It is one of the largest homogeneously-selected samples of
Seyfert galaxies available, and appears to contain representative populations
of galaxies with type 1 and type 2 optical spectra
(see also Tran 2001, Thean et al., 2001).
The selection was done at 12$\mu$m because this band contains a constant
fraction of bolometric flux for active galaxies ($\sim$ 1/7).
The 12$\mu$m selection is therefore the best
approximation so far available to a selection at a bolometric flux limit,
for the different types of active galaxies.
The resulting active galaxy sample contains 53
Seyfert type 1 galaxies (hereafter Seyfert 1's), 63
Seyfert type 2 galaxies (hereafter Seyfert 2's), two blazars and 38 high-luminosity
non-Seyfert galaxies out of a sample of 893 galaxies
(RMS).

Little is known about the FIR properties of type 1 higher redshift objects. Because of
their faintness in this energy domain they lack homogeneous samples and information
is still scant \cite{Hoo99,Wil99,Pol00}.
Type 2 AGN are much more difficult to find, and are missed in classical searches.
\hfill\break
In this work we make use of a homogenous data sample of optically selected type 1 AGN
\cite[]{And02} which was observed at far-IR wavelengths.
We make predictions of number counts and luminosities
that future FIR and submm space missions--
SIRTF (Space InfraRed Telescope Facility, NASA) and
the Herschel Space Observatory (Herschel, ESA)--will be able to detect.

For local galaxies we adopted the spectral energy distributions (SEDs)
of homogenous subsamples extracted from the 12$\mu$m galaxy sample:
Seyfert 1's, Seyfert 2's, starburst and normal galaxies
for which Infrared Space Observatory (ISO) photometry was collected
\cite[]{SAM}. For higher redshift type 1 AGN, we chose the
optically selected sample at a mean redshift of $<z>$=0.7 observed in the FIR
by Andreani et al. (2003).

\section{Expected counts of Type 1 AGN and Active Galaxies}

As it is well known, number counts of extragalactic objects are a
fundamental tool for investigating evolution, since they are related to
the Luminosity Function of the population, as a function of redshift. The
differential number counts, $dN(S)/dS$, (sources/unit flux interval/unit
solid angle) can be expressed as an integral of the
epoch-dependent luminosity function, $\Psi(L,z)$:

\begin{equation}
\frac{dN}{dS} = \int ^{z_u} _{z_l} dz \frac{dV}{dz} \frac{dL(S,z)}{dS}
\Psi [L(S,z),z]  ~~~~~~ sources/sr
\label{eq:diffcoun}
\end{equation}

\noindent
where $z_u$ and $z_l$ are the effective upper and lower limits of
the redshift distribution, and $\frac{dV}{dz}$ is the differential
comoving volume.
The integrated counts are found by integrating eq.\ref{eq:diffcoun}
to different flux limits, $S$.

To predict the differential and integral number counts of active galaxies
and quasars we need to compute equation \ref{eq:diffcoun} with the following
assumptions:

\begin{itemize}
\item
The geometry of the Universe: we use two different
models both with H$_0$=75 km/s/Mpc: (1) $\Omega=1$, $\Omega_\Lambda=0$ and
(2) $\Omega=1$, $\Omega_\Lambda=0.7$ and $\Omega_{mat}=0.3$.
They have been chosen because their predictions represent two extreme cases.
\item
The SEDs of the populations: we take the empirical optical-far-IR SEDs
of active galaxies and QSOs (see \S ~\ref{sec:SED}) and assume that they are
luminosity-independent. The only difference between normal and starburst
galaxies is the luminosity break. This simplified
assumption avoids more complicated situations
where the SED continuously changes with luminosity.
\item
The Local Luminosity Function (LLF): we consider different LLFs as empirically
determined for the different classes: AGN, starburst and normal galaxies (see \S~\ref{sec:LF}).
\item
The form of the LF evolution: different evolution laws are used for each
class (see \S ~\ref{sec:ev}).
\end{itemize}

In the following sections we discuss each of these assumptions.

\subsection{Far-IR SEDs of Quasars and Seyfert galaxies }
\label{sec:SED}

\begin{figure}[ht]
\figurenum{1} \vspace{10cm} \includegraphics{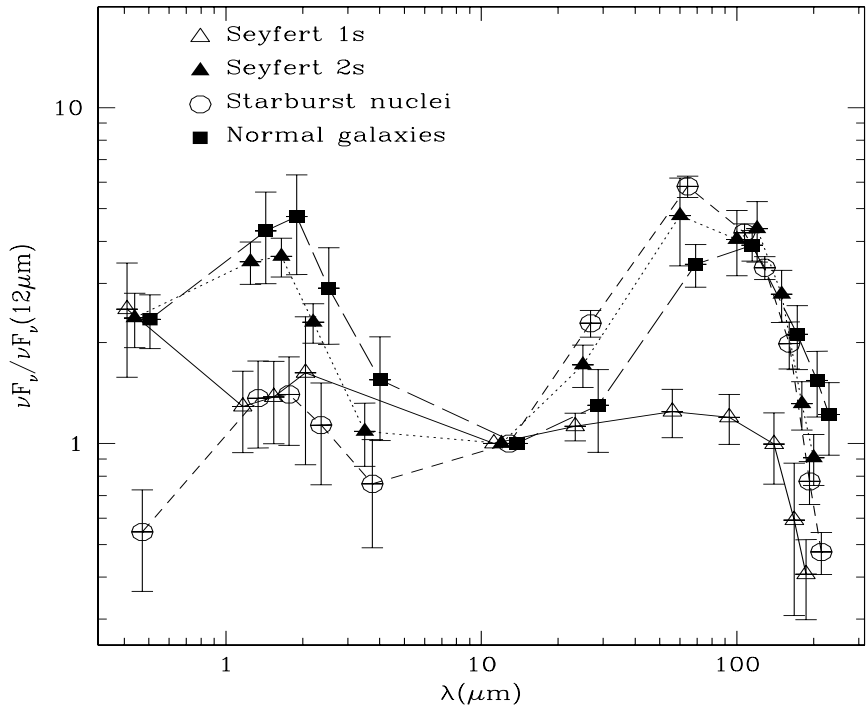} \vspace{1cm}
 \caption{\footnotesize
Average SEDs of Seyfert, Starburst and Normal Galaxies
normalized to 12$\mu$m. From Spinoglio, Andreani, Malkan (2002).}
\label{fig:agnsed}
\end{figure}
\begin{figure}[ht]
\figurenum{2} \vspace{8cm} \includegraphics{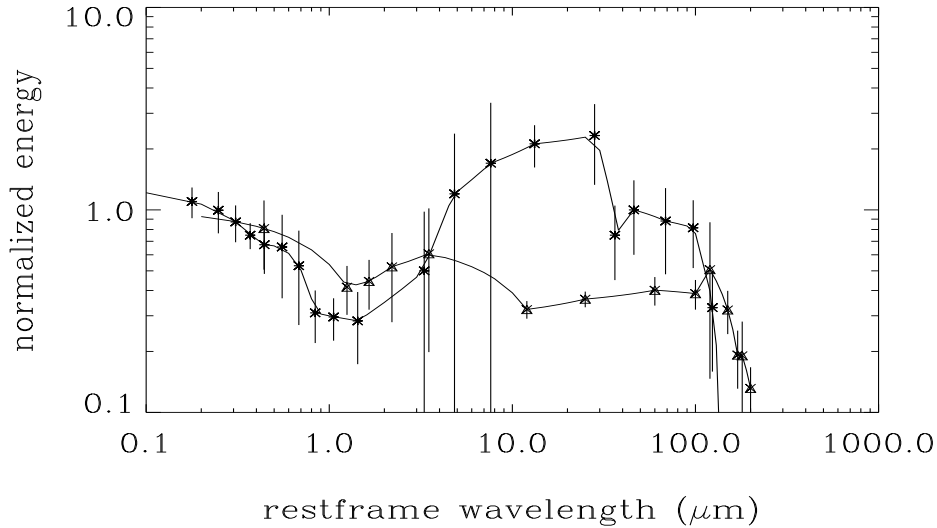} \vspace{1cm}
 \caption{\footnotesize
The average QSO spectrum (adopted from Andreani et al., 2003)
 compared with that of Seyfert
1's. Both spectra are normalized at $\lambda _{\rm blue} =0.44 \,\mu$m.
Triangles correspond to Seyfert 1 data, asterisks to QSOs.
Solid lines show the interpolating curves.
}
\label{fig:sedcomp}
\end{figure}

Based on the results of Spinoglio, Andreani and Malkan (2002),
Figure ~\ref{fig:agnsed} shows the observed average SED of the different
classes of galaxies in the Local Universe, normalized to 12$\mu$m.
Seyfert 1's have the flattest SED throughout the entire wavelength range,
with the relatively weakest FIR emission.
Seyfert 2's show two pronounced peaks: at 100$\mu$m and
1.6$\mu$m. Starburst galaxies show the brighter peak at 60$\mu$m and the weaker
one at 1.6$\mu$m, showing that recent and ongoing star formation--detected in
the FIR--outshines the old stellar populations, which peak
in the H band. Normal galaxies show again the same two peaks.

For this work we need to assume an
evolutionary law for AGN, but Seyfert 1's lack of a good empirical determination of
their redshift evolution. Fortunately, the evolution of
their higher redshift analogs--type 1 QSOs--is well determined.
Their far-IR SED is however less well
defined. In this work we take
the composite SED derived from optical, near-IR and FIR photometry
of an optically selected sample of QSOs.
Figure ~\ref{fig:sedcomp} shows the average spectrum in the QSO restframe,
normalized 
at $\lambda _{\rm blue} =0.44 \,\mu$m.
It has
a well defined IR component peaking at 10-30 $\mu m$ and dropping steeply
towards 100\,$\mu$m.
The errorbars shown are those related to the averages, but also include
photometric uncertainties.

Seyfert 1's and QSOs belong to the same class of objects, type
1 AGN, with Seyfert galaxies populating the low luminosity end of the
luminosity function. Figure ~\ref{fig:sedcomp} shows also a comparison of the
observed Seyfert 1's SED with that of QSOs.
Within the errorbars, no large differences are present, both below a few
microns and above 100\,$\mu$m. The QSO thermal spectrum
has a steeper turnover in the FIR range, and more power between 10 and
30\,$\mu$m compared to the Seyfert 1's SED.
QSO SED shows higher average dust temperature (100-150 K), with
respect to Seyfert 1's (22-70K) \cite[]{SAM}.
Because of their selection criteria, this difference could be ascribed to
a luminosity effect. We lack high-z object data to disentangle
this effect from an evolutionary behaviour.
\hfill\break
Note that the QSO SED is quite different from Seyfert 2's. This latter has
a large far-IR emission shifted to longer wavelengths, and a second bump in
the near-IR, which is not present in QSOs (see Figure \ref{fig:agnsed}).

To compute the K-corrections, i.e. $dL(S,z)/dS$ in equation \ref{eq:diffcoun},
we use for each class its appropriate composite spectrum as
shown in Figures~\ref{fig:agnsed} and ~\ref{fig:sedcomp}.

\subsection{The Luminosity Functions}
\label{sec:LF}

For the Local Luminosity Function (LLF), $\Psi_0(L)$, we adopt the 12$\mu$m
LLF for Seyfert, starburst and normal galaxies (RMS).
There is some uncertainty about the space densities
at the low luminosity end,
because no explicit correction for the over-concentration of local galaxies in
Virgo was applied.
Fang et al. (1998) and Xu et al. (1998) provide slightly different parameters,
but all 3 of those 12\,$\mu$m LLFs are in
reasonable agreement around L$_\star$ galaxies, which dominate the counts.
We use the same LLF shape with the different parameters given by RMS
for the different classes:
Seyfert 1's and 2's, and non-Seyferts (starburst and
normal galaxies).

This 12\,$\mu$m 'Seyfert 1' LF is in principle supposed to include
quasars in it, as it was computed for a sample that includes 3C 273.
As a practical limitation, though, it is too insensitive and thus covers too small
a volume to define the highest end of the Sy1/QSO LF. It also does not
take into account the AGN evolution which is, on the contrary,
included in the QSO LF.
Therefore, to extend the Sy1 LF to higher luminosities, and to include the
luminosity-number density evolution of these objects we use
the B-band luminosity function computed by Boyle et al. (2001) (see also
La Franca and Cristiani, 1998). We assume
that a constant relation exists between the B-band LF and the far-IR
one. This latter has been applied to both type 1 AGN (Seyfert 1's and QSO).

\begin{figure}[ht]
\figurenum{3} \vspace{8cm} \includegraphics{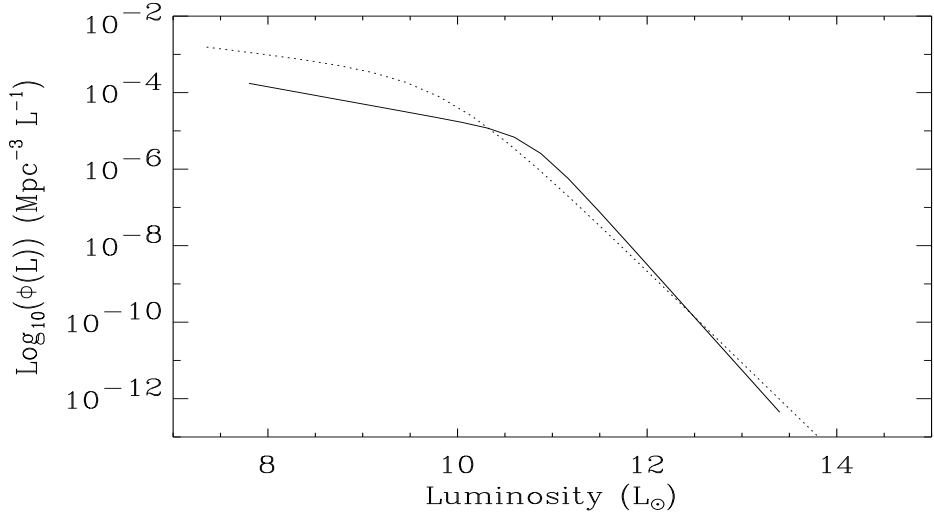} \vspace{1cm}
 \caption{\footnotesize
The QSO LF (solid line) computed at
a redshift of 0.3 is compared with the Seyfert 1 12\,$\mu$m LF (dotted line)
evolved in redshift as explained in section \ref{sec:ev}. The x-axis is
logarithm of object luminosity in solar luminosity.
}
\label{fig:LF}
\end{figure}

The Seyfert 1's and QSO used in this work were selected in different ways,
and the corresponding LF differs slightly in shape.
In Figure \ref{fig:LF} the QSO LF is shown (solid line)
superposed on that of the Seyfert 1's (dotted line).
Both LFs are computed at a redshift of 0.3, the minimun redshift for which a
reliable QSO LF exists. The Seyfert 1 LLF was evolved
from redshift 0 to a redshift of 0.3, according to the prescriptions
described in \S \ref{sec:ev}.

As shown in Figure \ref{fig:LF} the bright end of
the QSO LF is slightly steeper (the slope exponent being -2.5) than Seyfert 1's
(-2.2). However, within the errorbars this difference is not significant.
The knee at L$_\star$ differs and that of QSOs falls at higher luminosity
($\log$ L$_\star \sim$10.5).
At the faintest luminosities, the two LF slopes coincide but the
amplitudes differ by a factor of 10.
This high luminosity knee and the lower numbers at lower luminosities
for QSOs can be explained since nearly all QSO samples are based on integrated
properties of stellar-looking objects. When the
host galaxy starts becoming important, the object is no longer classified
as a QSO, but rather a Seyfert'1.
The best approximation of the type 1 luminosity function would be
to adopt (see Figure \ref{fig:LF}) the QSO LF above
$\log L_\star \sim 10.5$
and then the relatively steep Seyfert 1's below that, where they cross over.

Thus, for the sake of completeness we use, when computing the expected
number counts for Seyfert 1, two LFs: that of QSO
and that given by RMS for Seyfert 1.

\begin{table*}
\caption{Adopted SED, LLF and evolution for the different object classes}
\begin{tabular}{||c|cccc||}
\hline
\hline
Object type & SED & LLF & evolution & \\
\hline
QSO        & Figure~\ref{fig:sedcomp} & Bo01 & Bo01 + Bi01 &\\
Seyfert 1 & Figure~\ref{fig:agnsed} & Bo01 & Bo01 + Bi01 &\\
Seyfert 1 & Figure~\ref{fig:agnsed} & RMS (Seyferts 1) & MS01 &\\
Seyfert 2 & Figure~\ref{fig:agnsed} & RMS (Seyferts 2) & MS01 &\\
Starburst & Figure~\ref{fig:agnsed} & RMS (non-Seyfert) & MS01 &\\
Normal Galaxies & Figure~\ref{fig:agnsed} & RMS (non-Seyfert) & MS01 &\\
\hline
\hline
\end{tabular}
\tablenotetext{}{Bo01: Boyle et al. (2001)}
\tablenotetext{}{Bi01: Bianchi, Cristiani \& Tae-Sum (2001)}
\tablenotetext{}{RMS: Rush, Spinoglio \& Malkan (1993)}\\
\tablenotetext{}{MS01: Malkan \& Stecker (2001)}\\
\label{tab:parameters}
\end{table*}

\subsection{The Evolution } \label{sec:ev}

QSOs have a strong evolution, which is known at least
up to redshift 3. We adopt the redshift evolution of
the LF, $\Psi(L,z)$, described by Boyle et al. (2001) up to redshift 3, and
modified by Bianchi, Cristiani \& Tae-Sum (2001), whose model has a down-turn
at $z=3$ and an exponential decay until $z=10$.
For starburst and Seyfert galaxies we follow the prescriptions of Malkan and
Stecker (1998, 2001, hereafter MS01), who introduce a luminosity
evolution into the LLF:

\begin{equation}
\Psi(L(\nu),\nu;0)=C(\frac{L}{L_\star})^{1-\alpha}[1+(\frac{L}{\beta L_\star})^{-\beta}]
\frac{d\log \phi_{60}}{d\log \phi(\nu)}
\end{equation}

\noindent
where pure luminosity evolution is assumed, with luminosity scaling as
$(1+z)^Q$ up to redshift z$_{break}$, no further evolution for z$_{break}\leq$z
$\leq$z$_{max}$ and a cutoff at z$_{max}$=4.
We adopt the parameters of their {\it baseline} model:
Q=3.1 and z$_{break}$=2.
\hfill\break
Since Seyfert 1's can be considered as either low-luminosity type 1 AGN
or local type 1 active galaxies, we tried two possible
evolutionary laws for this class: the first one from MS01, the same used
for Seyfert 2's; the second one from Boyle et al. (2001), the same used for
QSOs. Table \ref{tab:parameters} summarizes the adopted laws.

\begin{figure*}[h]
\figurenum{4} \vspace{14cm} \includegraphics{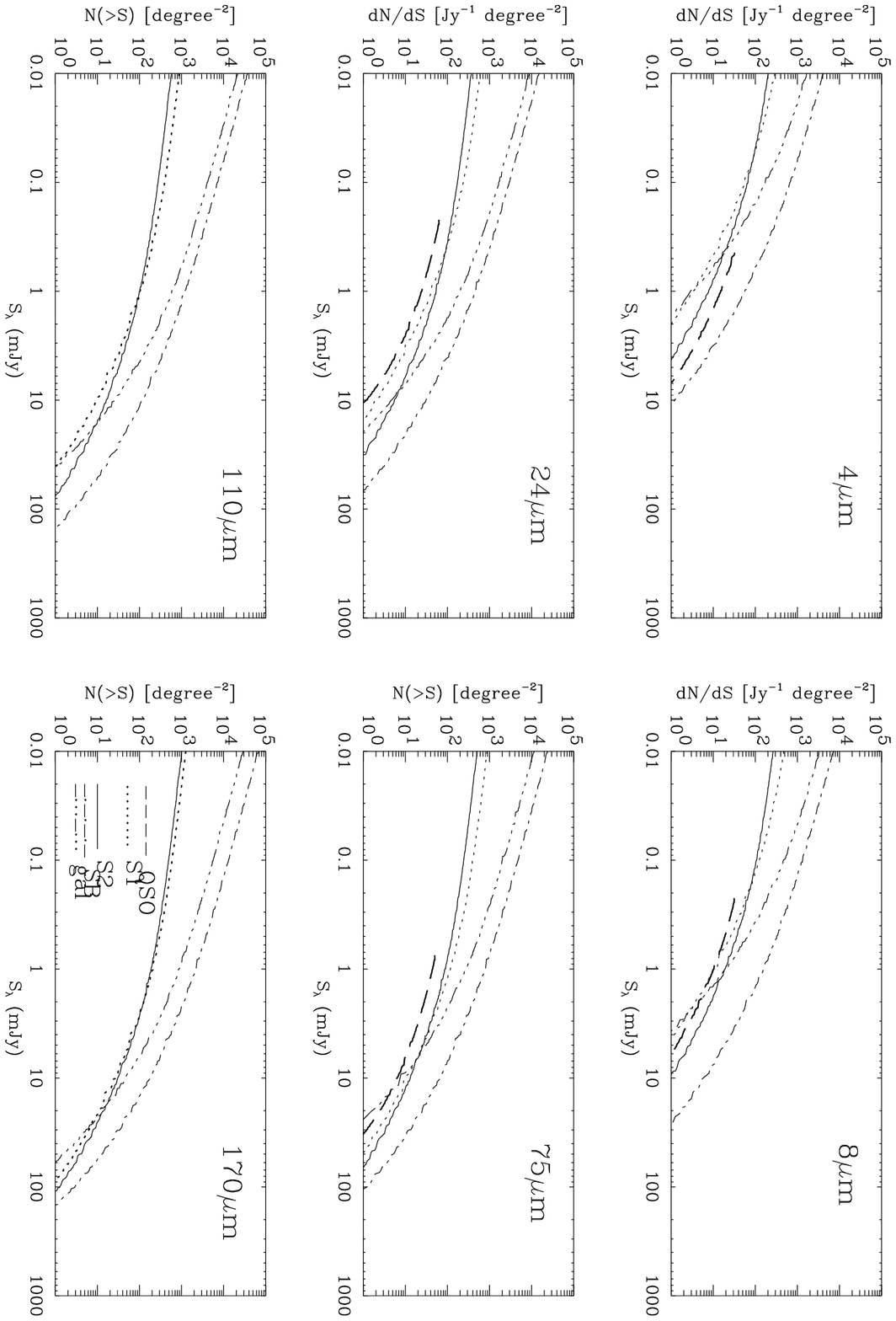} \vspace{1cm}
 \caption{\footnotesize
Expected differential number counts of type 1 QSO, Seyfert, starburst and
normal galaxies computed from the composite SEDs presented in this
work, for a Universe with $\Omega_\Lambda$=0, $\Omega_{\rm mat}=1$.
The different wavelengths correspond to the
bands of IRAC (4.5 and 8 $\mu$m) and MIPS (24$\mu$m) instruments on board
SIRTF and PACS (75, 110 and 170$\mu$m) on board Herschel.
}
\label{fig:counts}
\end{figure*}

Our definition of normal galaxies consists of all objects not showing any sign
of activity, below a luminosity cut-off of
${\rm L_{\rm IR}} < 6\times 10^{44}$ erg~s$^{-1}$ \cite{RMS}.
Most of them are then late-type
galaxies, whose evolution is assumed to be milder \cite[]{fra00}.
For this class
a different evolution law is then applied with a lower exponent, Q=2.

Recently it has been suggested that far-IR spectra of high redshift
galaxies do not show evidence for
any strong evolution with respect to local counterparts
(Chapman et al., 2002; Lagache, Dole, Puget, 2002).
Since at high redshifts we only detect the most luminous sources,
it is not currently possible to disentangle this finding from luminosity effects.
In other words, with the present data sets our
knowledge of how a galaxy goes through subsequent stages of
evolution is too scanty. Whether galaxy evolution is driven from
 number density or pure luminosity evolution cannot be settled.
\hfill\break
Because of our poor knowledge of galaxy evolution, we simply let
the population of normal galaxies have a milder evolution than starbursts.

We list for clarity in Table \ref{tab:parameters}
the adopted laws for SEDs, LLFs and evolution
for each class of objects.

\section{How many galaxies will SIRTF and Herschel detect ?}

Tables ~\ref{tab:counts1} and ~\ref{tab:counts2}
predict the counts in the filter bands of IRAC and MIPS on board
of SIRTF and PACS on board of Herschel.
Wavelengths and flux limits are given by the estimated instrument sensitivity
\footnote{from V\"ais\"anen (2001), SIRTF web-page: sirtf.caltech.edu and
PACS web-page: pacs.ster.kuleuven.ac.be}.
The exepcted survey sky area and counts for the different galaxy classes are shown.
Table ~\ref{tab:counts1} refers to $\Omega_\Lambda=0$ and
$\Omega=1$, Table ~\ref{tab:counts2} to $\Omega_\Lambda=0.7$
$\Omega_{\rm mat}=0.3$. In both cases H$_0$=75 km/s/Mpc.

Counts are computed by integrating down to the instrument sensitivity
over the given sky area.
We computed the counts in the bands at 110 and 170 $\mu$m
only for Seyfert, starburst and normal galaxies.
The empirical QSO SED
built on QSO photometry does not extend to these long wavelengths in the QSO
restframe. Consequently the counts cannot be computed unless
assumptions about the form of the long wavelength spectrum are made.

Tables ~\ref{tab:counts1} and ~\ref{tab:counts2}
clearly show that the number counts
strongly depend on Universe geometry. In a flat $\Lambda$-dominated Universe
the surface density of galaxies is larger. Open Cosmologies increase all the counts
by at least one order of magnitude in general. This effect is due to the
larger available comoving volume which
is only partially offset by the decrease in object number density in the
luminosity function (see eq.~\ref{eq:diffcoun}).
there is a larger comoving volume for a given surface area
on the sky, and therefore a larger surface density of high-$z$ objects.

Two different LFs and evolutionary laws are adopted for
Seyfert 1's, and the corresponding results are listed in two columns
in tables ~\ref{tab:counts1} and ~\ref{tab:counts2}.
Column 6 shows the predicted counts using the RMS local
LF as explained in section ~\ref{sec:LF} and evolved as
in section ~\ref{sec:ev}.
Column 5 shows instead the Seyfert 1 counts
computed using the QSO LF and integrating with an upper integration limit
corresponding to M$_{\rm B}$=-23 (as Seyfert 1's are considered complementary
to optical QSOs with respect to luminosity).
We find more 'Seyfert' galaxies than 'quasars' by one
order of magnitude at short wavelengths (4.5 and 8\,$\mu$m)
for both choices of the LFs, but their numbers coincide at 24 and 75\,$\mu$m.
Furthermore, quite similar numbers of Seyfert 1's are found (as listed in
columns 5 and 6) although
computed using two different LFs (see Figure \ref{fig:LF}).

\hfill\break
'Starbursts' outnumber the active galaxies at all wavelengths, but
the difference between the two populations shrinks as one goes from short
to long wavelengths.
This is as expected, since the normal
galaxy SED is  relatively brighter at the longest FIR wavelengths than
the starburst SED.

Malkan and Stecker (MS01) obtained similar results. They predict
914 galaxies in 0.3deg$^2$ (9.1 $\times 10^{-5}$ sr)
for the {\it baseline} case (Q=3.1 luminosity evolution index)
at 25\,$\mu$m, 5483 galaxies in 1 deg$^2$ (3$\times$10$^{-4}$ sr)
at 100\,$\mu$m 16,144 galaxies and at 200\,$\mu$m.
At these faint levels MS01's counts are dominated by SB galaxies, and are
only a little lower (roughly 30\% ) than the predicted counts in
tables ~\ref{tab:counts1} and ~\ref{tab:counts2}.
Also at 6\,$\mu$m MS01's prediction are a little lower, but not
seriously inconsistent.
The general conclusion is that the agreement between the two model
predictions of starburst and normal galaxies is satisfactory.

\begin{figure*}[h]
\figurenum{5} \vspace{14cm} \includegraphics{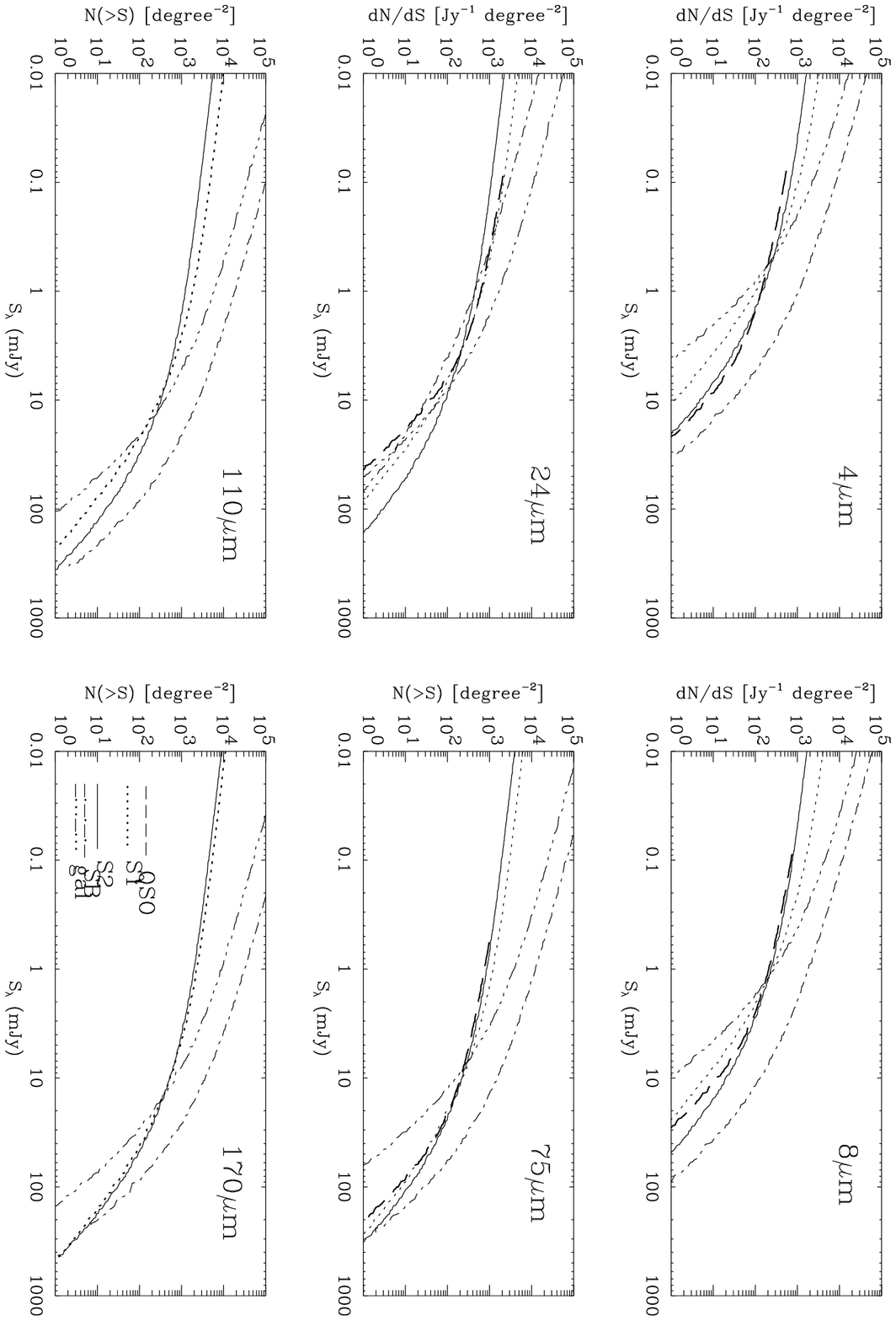} \vspace{1cm}
 \caption{\footnotesize
Same as Figure ~\ref{fig:counts}
but for a Universe with $\Omega_\Lambda=0.7$, $\Omega_{\rm mat}=0.3$}
\label{fig:diffcounts}
\end{figure*}

Figures \ref{fig:counts} and \ref{fig:diffcounts} plot
the expected differential counts for QSOs, Seyfert, starburst
and normal galaxies in the SIRTF observing bands: 4.5, 8, 24, 75, 110 and 170
\,$\mu$m
for the two Universe models considered in this work.
\hfill\break
The QSO counts break at fluxes around 0.1-1 mJy,
since the integration of equation ~\ref{eq:diffcoun} has a low
luminosity cut-off at M$_{\rm B}$=-23. This cut-off was introduced to
distinguish QSOs from Seyfert galaxies, and because the QSO LF is determined
down to this absolute blue magnitude.
This means that in future surveys {\it all} type 1 QSOs up to redshift 10
(those with M$_{\rm B} \le$ -23) will be detectable.

As expected, starbursts dominate the counts at all fluxes and all
wavelengths. Seyfert counts show a flatter slope
at low fluxes. Seyfert 1 counts grow more steeply than the Seyfert 2's
at low fluxes at almost every wavelength.\hfill\break
Seyfert 1's and 2's are almost indistinguishable at 170\,$\mu$m where
fluxes are very likely due to the host galaxies.
At $\lambda >$24\,$\mu$m Starbursts and normal galaxies show steeper slopes
at low fluxes.
Because of our assumptions on their milder evolution, the galaxy contribution
to counts is smaller at all wavelengths from a factor of 3 at low fluxes up
to two orders of magnitude at large fluxes.
At low fluxes and long wavelengths we expect a non-negligible
contribution from normal galaxies, although
starburts keep dominating the counts.
Starburst counts increase more steeply than Seyferts and behave as Seyfert 2's
at 4.5 and 8\,$\mu$m, except at low fluxes, where Seyfert 2's have a flatter slope.

\begin{table*}
\caption{Expected counts of AGN and galaxies in future IR
surveys in a $\Omega_\Lambda=0$, $\Omega=1$ Universe}
\begin{tabular}{||ccccccccc|c||}
\hline
\hline
$\lambda$ & S$_{\rm min}$(5$\sigma$,1h) & area & QSO & Sy 1 &
Sy 1 & Sy 2 & SB & gal & Instrument\\
($\mu$m) & ($\mu$Jy) & ($deg^2$) & & as QSO & as Sy 2 & & & &\\
\hline
4.5 & 13  & 0.3 & 5.0e2& 3.0e3 & 4.5e3 & 3.0e3&5.7e4 & 2.0e4& IRAC/SIRTF \\
8.0 & 98  & 0.3 & 6.4e2& 3.3e3 & 2.0e3 & 2.0e3&2.5e4 & 6.0e3& IRAC/SIRTF\\
24  & 550 & 0.3 & 1.8e3& 2.0e3 & 9.0e2 & 1.0e3&1.3e4 & 4.0e3& MIPS/SIRTF\\
75  & 3000& 1.0 & 4.0e3& 3.6e3 & 1.1e3 & 1.1e3&6.0e3 & 1.3e3& PACS/Herschel\\
110 & 3000& 1.0 & ...  & 3.0e3 & 1.3e3 & 1.6e3&1.2e4 & 3.6e3& PACS/Herschel\\
170 & 3000& 1.0 & ...  & 6.0e3 & 2.8e3 & 2.7e3&1.3e4 & 5.5e3& PACS/Herschel\\
\hline
\hline
\end{tabular}
\label{tab:counts1}
\end{table*}
\begin{table*}
\caption{Expected counts of AGN and galaxies in future IR
surveys in a $\Omega_\Lambda=0.7$, $\Omega_{\rm m}=0.3$ Universe}
\begin{tabular}{||ccccccccc|c||}
\hline
\hline
$\lambda$ & S$_{\rm min}$(5$\sigma$,1h) & area & QSO & Sy 1 &
Sy 1 & Sy 2 & SB & gal & Instrument\\
($\mu$m) & ($\mu$Jy) & ($deg^2$) & & as QSO & as Sy 2 & & & &\\
\hline
4.5 & 13  & 0.3 &2.0e4& 7.8e4 & 6.1e4 & 3.8e5 & 7.1e5 & 2.2e5 & IRAC/SIRTF \\
8.0 & 98  & 0.3 &2.6e4& 4.8e4 & 3.4e4 & 1.6e4 & 2.2e5 & 1.2e5 & IRAC/SIRTF\\
24  & 550 & 0.3 &8.0e4& 2.7e4 & 1.6e4 & 1.0e4 & 1.2e5 & 4.5e4 & MIPS/SIRTF\\
75  & 3000& 1.0 &2.0e5& 4.6e4 & 2.2e4 & 1.8e4 & 1.6e5 & 1.7e4 & PACS/Herschel\\
110 & 3000& 1.0 & ... & 4.7e4 & 2.7e4 & 2.7e4 & 2.1e5 & 3.5e4 & PACS/Herschel\\
170 & 3000& 1.0 & ... & 8.0e4 & 4.6e4 & 4.7e4 & 3.0e5 & 6.0e4 & PACS/Herschel\\
\hline
\hline
\end{tabular}
\label{tab:counts2}
\end{table*}

\section{What luminosities will future space missions detect?}

\begin{figure*}[ht]
\figurenum{6} \vspace{15cm} \includegraphics{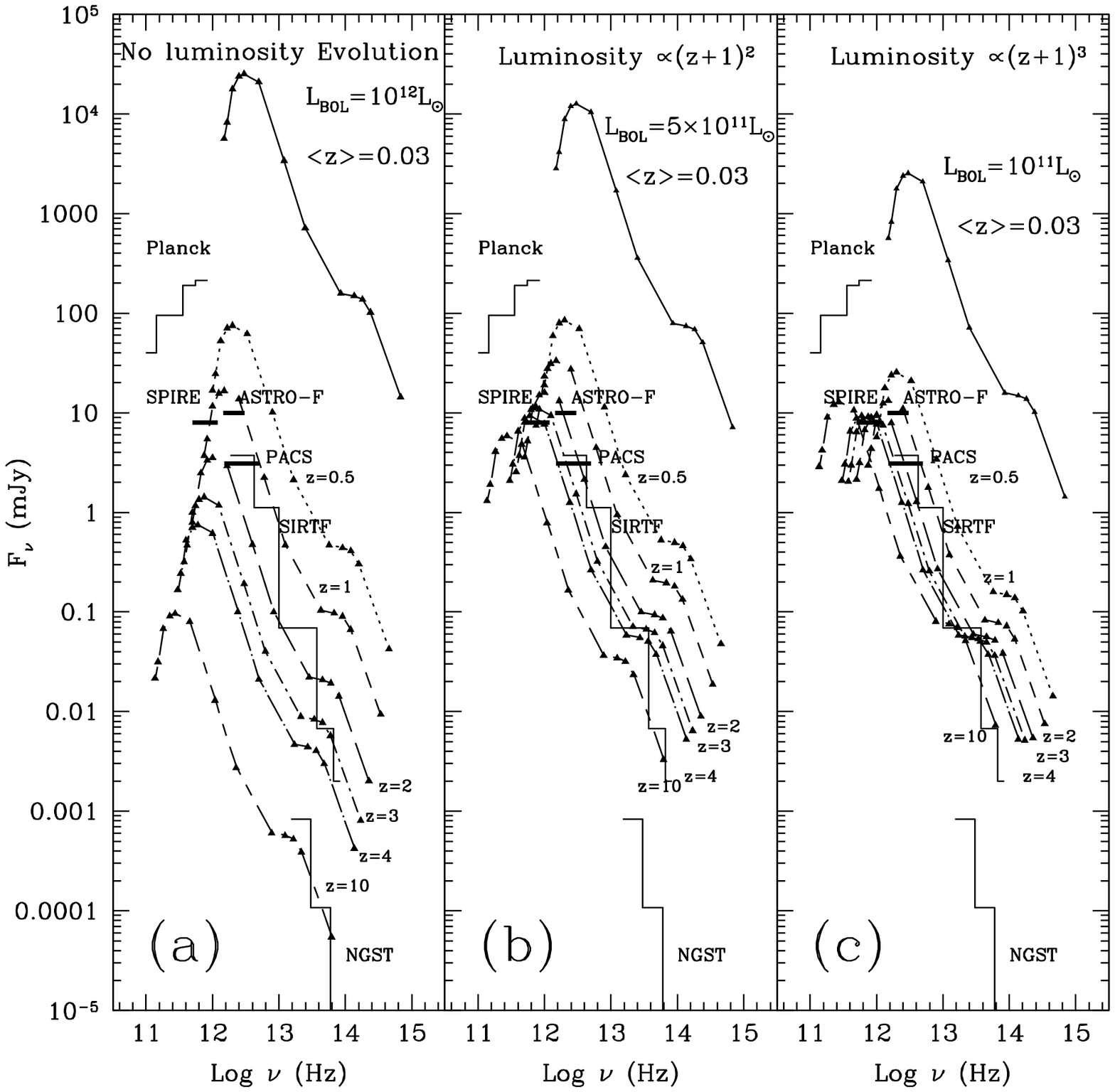} \vspace{1cm}
 \caption{\footnotesize
Predicted SEDs of starburst galaxies as a function of redshift, compared
to the expected sensitivities ( 5$\sigma$, 1 hr.) of the future space
missions Herschel, Planck, SIRTF, ASTRO-F and NGST with 3 different model
assumptions.
(a): SED of a galaxy with L$_{\rm BOL}=10^{12}$L$_\odot$, assuming no
luminosity evolution;
(b): SED of a galaxy with L$_{\rm BOL}=5\times 10^{11}$L$_\odot$, assuming
$(z+1)^2$ luminosity evolution;
(c): SED of a galaxy with L$_{\rm BOL}=10^{11}$L$_\odot$, assuming
$(z+1)^3$ luminosity evolution.
}
\label{fig:lum}
\end{figure*}

To explore the capabilities of future space missions we make the simple
exercise in Figure \ref{fig:lum} of comparing the expected sensitivities
of various instruments with the SEDs observed at redshifts 0 and 10.
We assume first
that the observed SED does not change with lookback time, and no evolution
occurs in luminosity (see panel (a)).
We then take two naive evolutionary models:
the first one with a luminosity evolution proportional to the second power of
the redshift (L $\propto$ (z + 1)$^{2}$) (see panel (b)) and the second one
proportional to the third power of the redshift (L $\propto$ (z +
1)$^{3}$) (see panel (c)). Figure \ref{fig:lum} shows the flux
distribution that a starburst galaxy of a given luminosity in the
Local Universe would have at different redshifts, assuming the
three different luminosity evolution laws. The predicted SEDs are
then compared with the expected sensitivities of
the future IR/submillimeter space missions, namely SIRTF, Planck,
Herschel, ASTRO-F and NGST. The first panel (Fig.\ref{fig:lum}a) shows that a
starburst galaxy with a bolometric luminosity of 10$^{12}$
L$_{\odot}$ can be detected (at 5 $\sigma$ in 1 hr) by SPIRE and
ASTRO-F at a redshift z=1; by PACS almost at a redshift z
$\simeq$ 2 at the longest wavelengths; by SIRTF at redshift
between 1 and 3, from long to short wavelengths, respectively; by
NGST at a hypothetical redshift close to 10. Similarly, the
second panel (Fig.\ref{fig:lum}b) shows that a starburst with a zero-redshift
bolometric luminosity of 5 $\times$ 10$^{11}$ L$_{\odot}$--
assuming a luminosity evolution proportional to the second power
of the redshift--would be detected by SPIRE on Herschel
at a redshift close to
z $\simeq$ 4; by ASTRO-F at z $\simeq$ 2; by PACS at z $\simeq$
3; by SIRTF at a redshift between 2 and 5, from long to short
wavelengths, respectively; by NGST at any redshift. Finally the
third panel (Fig. \ref{fig:lum}c) shows that a starburst with a zero-redshift
bolometric luminosity of 10$^{11}$ L$_{\odot}$--assuming a
luminosity evolution proportional to the third power of the
redshift--would be detected by SPIRE at a redshift greater than z
$\simeq$ 4; by ASTRO-F at z $\simeq$ 1; by PACS at z $\simeq$ 3;
by SIRTF at a redshift between 2 and an hypothetical z=10, from
long to short wavelengths, respectively; by NGST at any redshift.
The sensitivity of the all-sky survey that will be done with the
Planck mission will not be sufficient to detect statistically
significant numbers of galaxies beyond the Local Universe (unless
these are gravitationally lensed). Because starburst galaxies
have the steepest far-IR SED among the different classes and
- especially - much steeper than the Seyfert 1's, at the same
bolometric luminosity starburst galaxies can be more easily
detected by far-IR/submillimeter space missions, while Seyfert
1's are favoured by near- and mid-IR instruments.

We should keep in mind that
these L$_\star$ galaxies at z=4 will be swamped by hundreds of
times more galaxies at lower redshifts. The median redshifts of the
flux limited samples is 1.9, and around 3 percent of the total
number of active and starburst galaxies will be at redshift larger
than 3.

\section{Confusion Limits}

We now assess the implications of the above results for the confusion limits.
Confusion arises because of the uncertain and varying contribution of
flux density from the numerous unresolved faint sources that
fall within each resolution element (or beam in diffraction-limited instruments).
It depends on the details of both the shape of the counts and
the clustering strength of the galaxies in the survey. Confusion
becomes the dominant noise for any observations deeper than a certain
limit, which generally corresponds to a density of sources in the sky
that is greater than about 0.03/beam\footnote{the standard
rule-of-thumb for which confusion becomes important is at 1/30 of a source
per beam, a beam being one resolution element in the image}.

Confusion is a function of the beam size, $\theta _0$, the source counts,
$N(S)$, the cut-off related to the sources,
$x_c=S_c \cdot \Psi(\theta,\phi)$, the response distribution, $R(x)$,
and the $q$-parameter.
The cutoff $x_c$ is chosen usually to be $q$ times $\sqrt{<\delta i>^2}$
where

\begin{equation}
(\delta i)^2 = \int_0^{x_c} x^2 R(x) dx
\label{eq:conf}
\end{equation}

\noindent
and $ R(x) = \int \frac{dN(x/\Phi)}{dS} \frac{d\omega}{\Psi} $. The confusion
function can then be written (see i.e. Franceschini et al., 1989;
De Zotti et al., 1996; V\"ais\"anen et al., 2001):

\begin{equation}
\sigma _{conf} = f (\theta _0, N(S), x_c, S_c, q)
\end{equation}

\noindent
If differential counts follow a power-law behaviour, $\frac{dN(S)}{dS}
\propto k S^{-(\delta+1)}$ then eq.~\ref{eq:conf} simplifies to
$\sigma _{conf} = (\frac{q^{2-\delta}}{2-\delta})^{1/\delta}
(k\Omega_e)^{1/\delta}
$
\noindent
where $\Omega_e = \int [\Psi (\theta,\phi)]^\delta d\omega$] is the
effective beam and $\Psi$ the angular power pattern of the instrument.

In order to estimate the value of $\sigma _{conf}$ at each wavelength, we sum up
the contribution of each source population to the counts to integrate
eq.~\ref{eq:conf}. The resulting values are shown in table \ref{tab:conf}.
\hfill\break
Although the adopted Universe models are quite extreme, and show
a huge discrepancy in predicting source counts (see tables
~\ref{tab:counts1} and ~\ref{tab:counts2})
the values reported in Table \ref{tab:conf}
can be considered as a reasonable range of possible limiting fluxes
which should be considered when planning deep integrations.
\hfill\break
The values are not in conflict with the sensitivities shown in tables
~\ref{tab:counts1} and ~\ref{tab:counts2}. They represent very conservative
limits due to source confusion which depends on the models
we adopted. They mean, for instance, that IRAC at 4.5\,$\mu$m will never be
confusion limited for integration times shorter than 700 hours in a flat
Universe with $\Omega_\Lambda=0$, while after 2 hours
its performance will be limited by source confusion in an open Universe
with $\Omega_\Lambda>0$. PACS at 75\,$\mu$m will never be confusioned limited
in both cases.
At longer wavelengths the PACS instrument will always be confusion limited
for integration time longer than 1 hour at 110\,$\mu$m
and 20 m at 170\,$\mu$m, while MIPS at 24\,$\mu$m reaches
confusion limits in a $\Omega_\Lambda>0$ Universe for integrations longer than
nine hours.

\hfill\break
Although the results of these computations are affected by uncertainties --
linked to source evolution and Universe geometry --,
they show that
confusion will constitute a serious fundamental detection limit in these
forthcoming experiments. Overcoming this limit will eventually require
interferrometric observations, such as those anticipated at
350--450\,$\mu$m by ALMA.

\begin{table*}
\caption{Expected 5$\sigma$ confusion limits for future instruments}
\begin{tabular}{||cccc|c||}
\hline
\hline
$\lambda$ & $\Omega_e$ & $\sigma_{conf}$($\Omega_\Lambda=0,
\Omega=1$) &$\sigma_{conf}$($\Omega_\Lambda=0.7,\Omega_{\rm mat}=0.3$)
& Instrument\\
($\mu$m)  & (10$^{-5}$ sr) & ($\mu$Jy) & ($\mu$Jy) & \\
\hline
4.5 & .64 & 0.5   &  4     & IRAC/SIRTF \\
8.0 & 1.2 & 3     & 30     & IRAC/SIRTF\\
24  & 3.4 & 30    & 200  & MIPS/SIRTF\\
75  & 2.4 & 30    & 600   & PACS/Herschel\\
110 & 2.9 & 100   & 3000  & PACS/Herschel\\
170 & 4.5 & 600   & 20000 & PACS/Herschel\\
\hline
\hline
\end{tabular}
\label{tab:conf}.
\end{table*}

\section{Colour-colour diagrams}
\label{sec:colcol}

\begin{figure*}[ht]
\figurenum{7} \vspace{13cm} \includegraphics{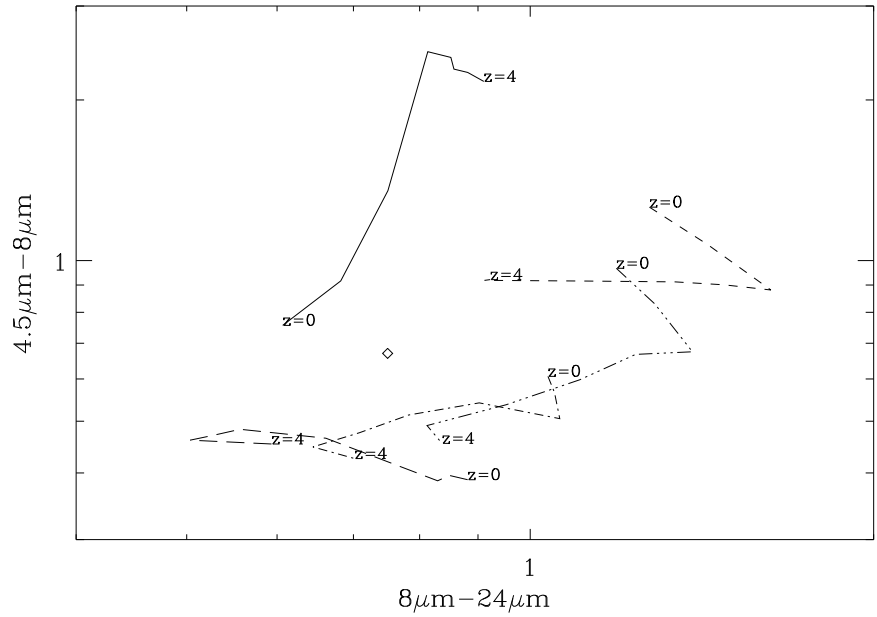} \vspace{1cm}
\caption{\footnotesize
Redshift tracks for QSO (solid line), Seyfert 1 (dash), Seyfert 2
(dash-dot), starburst (long dash ) normal (dash dot dot dot)
galaxies from z=0 to z=4.
The diamond refers to the most extreme measured values
of these colours for class I protostars (Nielbock et al. 2001).}
\label{fig:colcol1}
\end{figure*}
\begin{figure*}[ht]
\figurenum{8} \vspace{13cm} \includegraphics{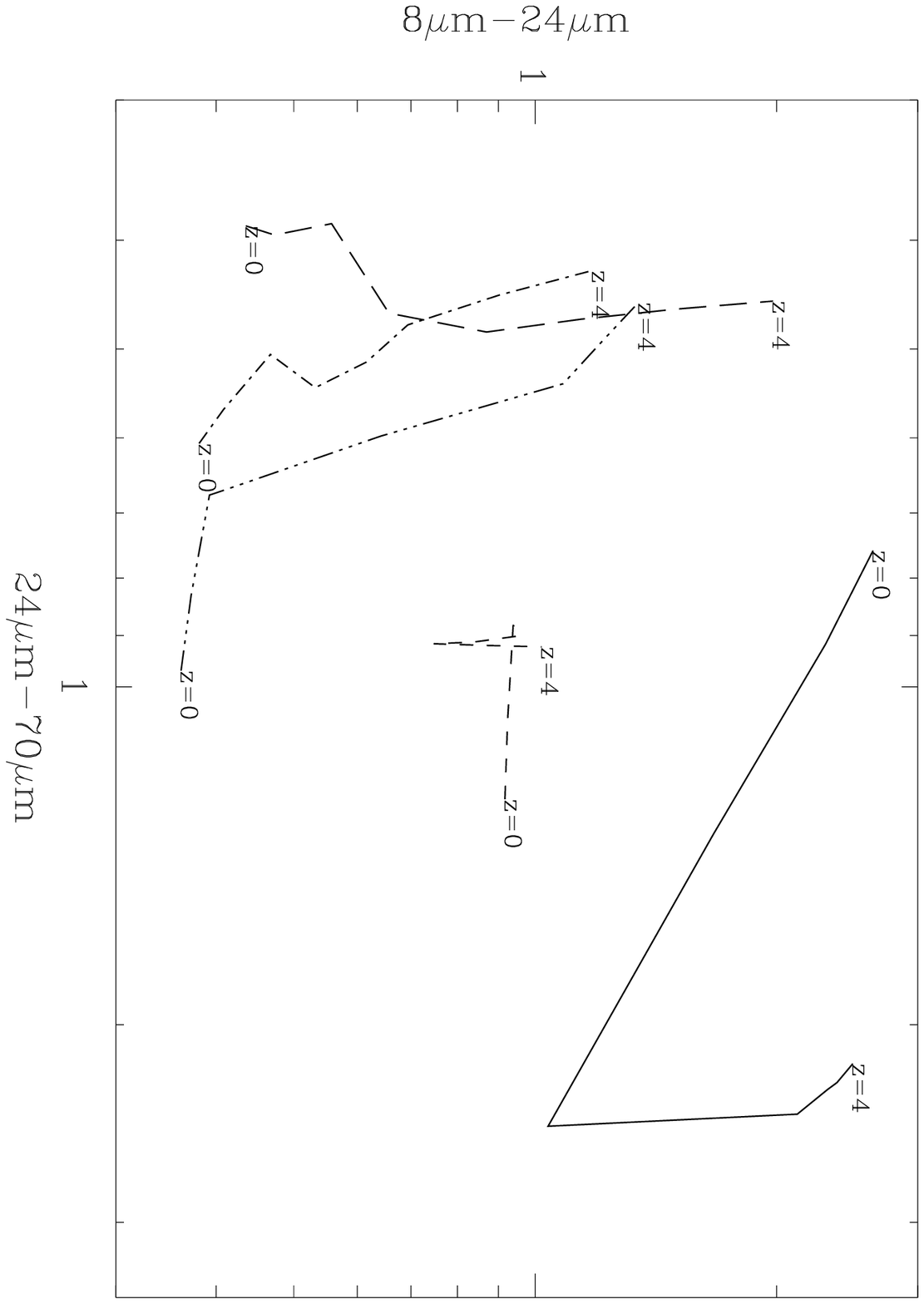} \vspace{1cm}
\caption{\footnotesize
Redshift tracks for QSO (solid line), Seyfert 1 (dash), Seyfert 2
(dash-dot), starburst (long dash ) normal (dash dot dot dot)
galaxies from z=0 to z=4. }
\label{fig:colcol2}
\end{figure*}

Colour-colour diagrams of objects observed
in multifrequency surveys are a powerful tool for
identifying and classifying objects. 
In
Figures \ref{fig:colcol1} and \ref{fig:colcol2} the colour-colour diagrams
[4.5 - 8] versus [8 - 24] and [8 - 24] versus [24-70]
\footnote{we use as colour definition: $[\lambda_1 - \lambda_2] = d\log\frac{
(F_1/F_2)}{\lambda_1/\lambda_2}$, where $F_i$ are flux densities.} are presented.
In these two diagrams redshift tracks, computed from redshifts of 0 to 4,
are shown for each type of galaxy.
These tracks
identify the locus on the diagram where object colours are expected to lie,
and can be then exploited to pinpoint candidates for follow-up
observations.

To avoid confusion with galactic stellar objects, we also plot in
Figure \ref{fig:colcol1} the average colours of Class I protostars
\cite[]{Nie01}, while Class 0 protostars
\cite[]{And00}, and brown dwarfs \cite[]{Mar01} are, respectively, too
red and too blue in at least one colour
to fall within the limits of this diagram. The diagrams in
Figures \ref{fig:colcol2} are never affected by the presence of stellar objects.

In the [4.5 - 8] vs [8 - 24] diagram (Figure \ref{fig:colcol1})
a clear separation occurs among the three
classes of galaxies: quasars, Seyfert 1's and starburst galaxies. Their
tracks are constrained by: $0.6<[8 - 24]<0.9$ and $[4.5 - 8]>0.8$,
$[8 - 24]>1.0$ and $[4.5 - 8]>0.9$, $0.5<[8 - 24]<0.9$ and
$[4.5 - 8]<0.5$, respectively.
However, both Seyfert 2's and normal galaxies lie in a region which
is intermediate between Seyfert 1's and starburst galaxies. Therefore it
would not be possible to
distinguish them solely on the basis of 4.5-8-24$\mu$m photometry.

The [8 - 24] vs [24 - 70] diagram (Figure \ref{fig:colcol2})
shows that quasars and Seyfert 1's are both well
segregated from starbursts, Seyfert 2's and normal galaxies,
and can be selected by [8 - 24]$>$0.9 and [24 - 70]$>$0.8.
However, these three latter galaxy classes
do overlap each other in these colours.
Starbursts and Seyfert 2's are almost indistinguishable in this diagram,
and normal galaxies share most of their colours except at small redshifts
where [24 - 70]$>$0.7.

In the diagrams in Figures \ref{fig:colcol1} and \ref{fig:colcol2}
the average local SED of each type of galaxies is
considered, assuming no scatter.
To have a better estimate of the power of the two colour-colour
diagrams in separating different classes of objects and of their limits,
we have estimated the likely errors in the colours and included them in the figures,
assuming that the flux densities
in the three bands have uncertainties of 20\%.

These colour-colour plots are affected by a redshift degeneracy:
higher-redshift Seyfert 2's resemble the lower-redshift
starburst galaxies. For most objects the redshift will not be known.
Using more than just 3 filters simultaneously will help in sorting
this out.
In fact, some tracks are smooth enough that it is possible to fit a simple
analytic formula for each type of galaxy:

\begin{equation}
z_{\rm object ~type} = A [4-8] + B [8-24] + C [24 - 70] + D [4-8]^2 + E
[8-24]^2 + F [24 - 70]^2
\end{equation}

\noindent
with coefficients given in Table ~\ref{fit_coef}.

\begin{table*}
\caption{Fitting coefficients for photometric redshift estimates}
\begin{tabular}{||c|cccccc||}
\hline
\hline
Object type & A & B & C & D & E & F \\
\hline
starburst       & .096 & 1.88  &  .05 &  .73 &  .41 & .84 \\
Seyfert 1's     & .310 & 1.85  & 3.03 & -.69 & -.95 & .22 \\
Seyfert 2's     & .990 & 1.95  & 0.31 & -.13 &  .26 & .85 \\
normal galaxies & .460 & 1.42  & 0.74 & -.56 &  .43 & .64 \\
Quasars         &-.160 & 1.99  &-0.05 & -1.14 & .40 & .34\\
\hline
\hline
\end{tabular}
\label{fit_coef}.
\end{table*}

Evolution from one class of galaxies to another has not been
considered, as well as other speculative assumptions, because we want
to keep this work as much as possible on a firm empirical basis.

\section{Conclusions}

\begin{itemize}
\item
In this paper we predict, on an empirical basis,
active and normal galaxy counts which future space missions
will be able to measure in the far-IR.
The observed SEDs of different types of objects
(Seyfert 1 and 2, starburt, normal galaxies and QSOs)
are evolved backwards in time, using our present knowledge of their local
luminosity functions and their evolution.
\item
The number of Seyfert 1 and 2, Starbursts and normal galaxies
that should be detected in future surveys are shown in Tables
~\ref{tab:counts1} and ~\ref{tab:counts2} and Figures \ref{fig:counts}
and \ref{fig:diffcounts}. In the most conservative case (a flat Universe
$\Omega_\Lambda=0$, $\Omega=1$), SIRTF and Herschel will be able to detect from
thousands to tens of thousands of active and starburst galaxies in a field
of 0.3 and 1 deg$^2$, respectively.
Our results are affected by the
following uncertainties:
\begin{itemize}
\item
Type 1 QSOs: their evolution has a solid foundation since
the B-band LF is well
known at least up to redshift 3. Their far-IR SED is however very uncertain.
We use the composite spectrum shown
in Figure~\ref{fig:sedcomp} whose photometric uncertainty is large (see
Andreani et al., 2003 for details).
\item
Nearby Seyfert and Starburst galaxies on the contrary have a well-defined
average
SED, but the knowledge of their evolution is poor.
Future far-IR Space missions,
mainly SIRTF, will eliminate this uncertainties.
\end{itemize}
\item
Colour-colour diagrams are constructed from the SEDs shown in
Figures \ref{fig:sedcomp} and \ref{fig:agnsed}, without evolution taken into
account. Curves are
very uncertain for QSOs, because of their poorly determined far-IR SED.
Colour-colour diagrams with the selected frequency bands 
will be a powerful tool for discriminating among different object types:
quasars from Seyfert 1's,
Seyfert 2's and starburst galaxies. Redshift measurements (or estimates) can be
used to resolve the degeneracy present in some cases.

\item
The confusion-limiting flux tends to be high
compared with the
limiting sensitivity of both SIRTF and Herschel instruments. Instrument teams are
indeed seriously considering it for very long exposures.
\item
SIRTF will make fundamental discoveries about AGN evolution,
exploiting mainly the 8 and 24\,$\mu$m bands. These bands are critical to selecting
unbiased samples of type 1 and 2 AGN.
\end{itemize}

\acknowledgements
PA acknowledges the Infrared Group of the Max-Planck Institut f\"ur
Extraterrestrische Physik of Garching (Germany)
for hospitality during the first stage of this work.
The Italian Space Agency (ASI) has funded part of this work under the contract
ASI-I-R-037-01.

\end{document}